\documentclass[runningheads]{llncs}
\usepackage{graphicx}
\usepackage{appendix}
\usepackage{tikz}
\usepackage{comment}
\usepackage{amsmath,amssymb} 
\usepackage{color}
\usepackage{tikz}
\usepackage{bbding}
\usepackage{pifont}
\usepackage{multirow}

\newcommand{\xmark}{\ding{55}}

\usepackage[accsupp]{axessibility}  

\begin{document}
\pagestyle{headings}
\mainmatter
\def\ECCVSubNumber{7}  

\title{When CNN Meet with ViT: \\
Towards Semi-Supervised Learning for Multi-Class Medical Image Semantic Segmentation} 


\titlerunning{CNN and ViT for Semi-Supervised Medical Image Segmentation}
%
\author{Ziyang Wang\inst{1} \and
Tianze Li\inst{2} \and
Jian-Qing Zheng\inst{1} \and
Baoru Huang\inst{2}
}
\authorrunning{Z. Wang et al.}
%
\institute{University of Oxford, UK  \and
Imperial College London, UK \\
\email{ziyang.wang@cs.ox.ac.uk}
}

\maketitle

\begin{abstract}

Due to the lack of quality annotation in medical imaging community, semi-supervised learning methods are highly valued in image semantic segmentation tasks. In this paper, an advanced consistency-aware pseudo-label-based self-ensembling approach is presented to fully utilize the power of Vision Transformer(ViT) and Convolutional Neural Network(CNN) in semi-supervised learning. Our proposed framework consists of a feature-learning module which is enhanced by ViT and CNN mutually, and a guidance module which is robust for consistency-aware purposes. The pseudo labels are inferred and utilized recurrently and separately by views of CNN and ViT in the feature-learning module to expand the data set and are beneficial to each other. Meanwhile, a perturbation scheme is designed for the feature-learning module, and averaging network weight is utilized to develop the guidance module. By doing so, the framework combines the feature-learning strength of CNN and ViT, strengthens the performance via dual-view co-training, and enables consistency-aware supervision in a semi-supervised manner. A topological exploration of all alternative supervision modes with CNN and ViT are detailed validated, demonstrating the most promising performance and specific setting of our method on semi-supervised medical image segmentation tasks. Experimental results show that the proposed method achieves state-of-the-art performance on a public benchmark data set with a variety of metrics. The code is publicly available.\footnote[1]{https://github.com/ziyangwang007/CV-SSL-MIS}

\end{abstract}

\section{Introduction}

\begin{figure*}[ht!]  
\centering  
\includegraphics[width=0.95\linewidth]{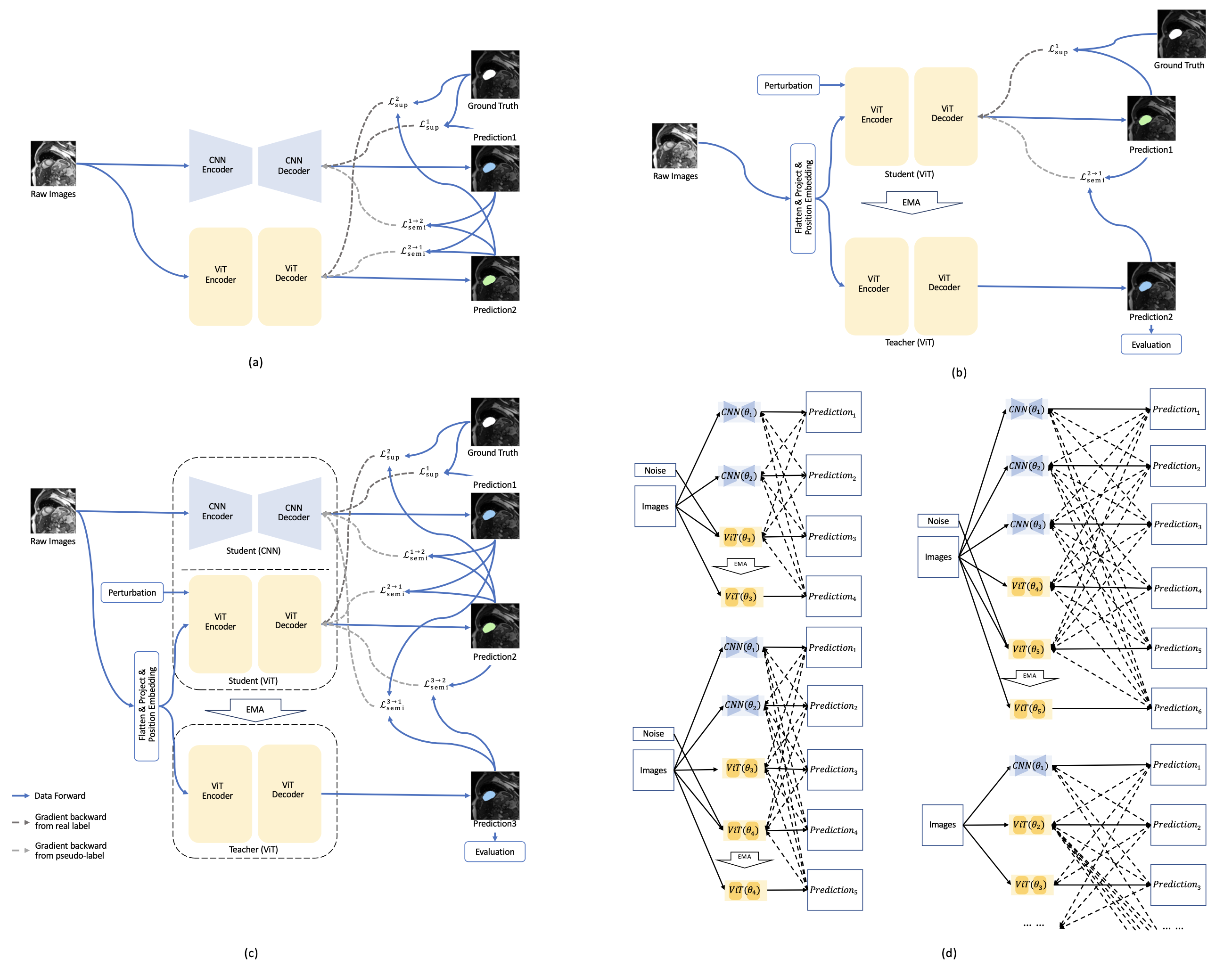}  
\caption{The Example 2-Model-Based, 3-Model-Based, and 4-Model-Based SSL Framework for Image Segmentation. The supervision mechanism is illustrated by minimizing the difference (also known as $Loss$) between prediction and (pseudo) label. (a) The best 2-model-based SSL framework~\cite{luo2021semi}. (b) The pure ViT-based student-teacher style SSL framework. (c) The best 3-model-based SSL framework, i.e. S4CVnet. (d) The 4-model-based, 5-model-based SSL framework. }  
\label{fig:sscvnet}  
\end{figure*}

Medical image segmentation is an essential task in computer vision and medical image analysis community where deep learning methods have shown dominated position recently. The promising results of current deep learning study not only relies on architecture engineering of CNN~\cite{long2015fully,ronneberger2015u,chen2018encoder,wang2020deep}, but also on sufficient high-quality annotation of data set~\cite{bernard2018deep,dosovitskiy2020image,wang2022uncertainty,deng2009imagenet}. The most common situation of clinical medical image data, however, is with a small amount of labelled data and a large number of raw images such as CT, ultrasound, MRI, and videos from laparoscopic surgery~\cite{luo2021semi,huang2022simultaneous,wang2021rar,wang2022icip}. In recent studies of neural network architecture engineering, the performance of purely self-attention-based, Transformer~\cite{vaswani2017attention}, outperforms CNN and RNN because of the ability of modeling long-range dependencies~\cite{liu2021swin,dosovitskiy2020image}.
Following the above concern of data situation, and the recent success in network architecture engineering, we hereby proposed a {\bf S}emi-{\bf S}upervised medical image {\bf S}emantic {\bf S}egmentation framework aiming to fully utilize the power of {\bf C}NN and {\bf V}iT simultaneously, called {\bf S4CVnet}. The framework of S4CVnet consists of a feature-learning module, and a guidance module, which is briefly sketched in Figure~\ref{fig:sscvnet}. This setting is inspired by the Student-Teacher style framework~\cite{han2018co,tarvainen2017mean,wang2022uncertainty}, that the perturbation is applied to the student network, and the parameters of the teacher network are updated through Exponential Moving Average(EMA)~\cite{laine2016temporal} which makes the teacher network much robust to guide the learning of the student network with pseudo label under consistency-aware concern. To utilize the feature-learning power of CNN and ViT simultaneously and avoid the barrier caused by the different architecture of two networks, we hereby come up with a dual-view co-training approach in the feature-learning module\cite{wang2022icip,chen2021semi}. Two different views of networks infer pseudo labels simultaneously to expand the size of the data set with raw data, complementing and beneficial to each other during the training process. One feature-learning network is also considered a bridge to be applied network perturbation and transfer of learning knowledge via the Student-Teacher style scheme.

The contributions of S4CVnet is fourfold and discussed as follows:
\begin{itemize}
    \item an enhanced dual-view co-training module aiming to fully utilize the feature-learning power of CNN and ViT mutually is proposed. Both CNN and ViT are with the same U-shape Encoder-Decoder style segmentation network for fair comparison and exploration, 
    \item a robust guidance module based on computational efficient U-shape ViT is proposed, and a consistency-aware Student-Teacher style approach via EMA is properly designed,
    \item an advanced semi-supervised multi-class medical image semantic segmentation framework is proposed, evaluated on a public benchmark data set with a variety of evaluation measures, and keeps state-of-the-art against other semi-supervised methods under the same setting and feature information distribution to our best of knowledge~\cite{tarvainen2017mean,zhang2017deep,verma2019interpolation,vu2019advent,yu2019uncertainty,wang2022triple,qiao2018deep,luo2021semi,wang2022icip,wang2022uncertainty},
    \item a topological exploration study of all alternative supervision modes with CNN and ViT, as well as an ablation study, is validated to present a whole picture of utilizing CNN and ViT in a semi-supervised manner, and demonstrates the most proper setting and promising performance of S4CVnet.
\end{itemize}

\section{Related Work}

{\bf Semantic Segmentation} The convolutional neural network(CNN) for image semantic segmentation, as a dense prediction task, has been widely studied since 2015, i.e. FCN~\cite{long2015fully}. It is the first CNN-based network trained with a supervised fashion for pixels-to-pixels prediction tasks. Then, the subsequent study of segmentation was dominated by CNN with three aspects of contribution: backbone network, network blocks, and training strategy. For example, one of the most promising backbone networks is UNet~\cite{ronneberger2015u}, which is an Encoder-Decoder style network with skip connections to efficiently transfer multi-scale semantic information. A variety of advanced network blocks to further improve CNN performance such as attention mechanism~\cite{woo2018cbam,jaderberg2015spatial}, residual learning~\cite{he2016deep}, densely connected~\cite{huang2017densely}, dilated CNN~\cite{chen2017deeplab} have been applied to the backbone network, UNet, which results in a family of UNet~\cite{ibtehaz2020multiresunet,wang2021rar,wang2021quduraple,isensee2018nnu}. The CNN for dense prediction tasks, however, is lack of ability of modelling long-range dependencies in recent studies, and is defeated by Transformer, a purely self-attention-based network, that originated from natural language processing~\cite{vaswani2017attention}. The Transformer was widely explored in computer vision tasks, i.e. Vision Transformer(ViT)~\cite{dosovitskiy2020image}, around classification, detection, and segmentation tasks~\cite{liu2021swin,strudel2021segmenter,chen2021transunet,cao2021swin}. In this paper on the backbone aspect, we focus on exploring the feature-learning power of CNN and ViT simultaneously, enabling both of them beneficial to each other, and specifically tackling a semi-supervised dense prediction task based on a multi-view co-training self-ensembling approach. \\
{\bf Semi-Supervised Semantic Segmentation} Besides the study of backbone networks and network blocks, the training strategy is also an essential study depending on the different scenarios of data set, such as weakly-supervised learning to tackle low-quality annotations~\cite{zhou2016learning,chang2020weakly,song2019box}, noisy annotations~\cite{wang2021rar}, multi-rater annotations~\cite{ji2021learning}, and mixed-supervised learning for multi-quality annotations~\cite{reiss2021every}. The most common situation of medical imaging data is with a small amount of labelled data and a large amount of raw data due to the high labelling cost, so semi-supervised learning is significantly valuable to be explored. Co-training, and self-training are two widely studied approaches in semi-supervised learning. Self-training, also known as self-labelling, is to initialize a segmentation network with labelled data at first. Then the pseudo segmentation masks on unlabelled data are generated by the segmentation network~\cite{you2011segmentation,chen2020naive,ibrahim2020semi,zoph2020rethinking,mendel2020semi,luo2021urpc}. A condition is set for the selection of pseudo segmentation masks, and the segmentation network is retrained by expanding training data several times. GAN-based approaches mainly studied how to set the condition using discriminator learning for distinguishing the predictions and the ground-truth segmentation~\cite{hung2018adversarial,souly2017semi}. The other approach is Co-training, which is usually to train two separate networks as two views. These two networks thus expand the size of training data and complement each other. Deep Co-training was firstly proposed in~\cite{qiao2018deep} pointing out the challenge of utilizing co-training with a single data set, i.e. `collapsed neural networks'. Training two networks on the same data set cannot enable multi-view feature learning because two networks will necessarily end up similar. Disagreement-based Tri-net was proposed with three views which improved with diversity augmentation for pseudo label editing to solve `collapsed neural networks' by 'View Differences'~\cite{dong2018tri,wang2022triple}. Uncertainty estimation is also an approach to enable reliable pseudo labels to be utilized to train other views~\cite{xia20203d,wang2022uncertainty,chen2021semi}. Current key studies of Co-training mainly on: (a)enabling the diversity of two views, and (b)properly/confidently generating pseudo labels for retraining networks. In this paper of the training strategy aspect, we adopt two completely different segmentation networks to encourage the difference between two views in the feature-learning module. Furthermore, inspired by the Student-Teacher style approach~\cite{tarvainen2017mean,laine2016temporal}, a ViT-based guidance module is developed which is much more robust with the help of the feature-learning module via perturbation, and average model weights~\cite{laine2016temporal,gal2016dropout}. The guidance module is able to confidently and properly supervise the two networks of the feature-learning module in the whole semi-supervision process via pseudo label.

\section{Methodology}

In generic semi-supervised learning for image segmentation tasks, $\mathbf{L}$, $\mathbf{U}$ and $\mathbf{T}$ normally denote a small number of labelled data, a large amount of unlabeled data, and a testing data set. We denote a batch of labeled data as $(\textbf{\textit{X}}_{\rm l}, \textbf{\textit{Y}}_{\rm gt}) \in \mathbf{L}$, $(\textbf{\textit{X}}_{\rm t}, \textbf{\textit{Y}}_{\rm gt}) \in \mathbf{T}$ for labeled training and testing data with its corresponding ground truth, and a batch of only raw data as $(\textbf{\textit{X}}_{\rm u})\in \mathbf{U}$ in the unlabeled data set, where $\textbf{\textit{X}} \in \mathbb{R}^ {h \times w} $ representing a 2D gray-scale image. $\textbf{\textit{Y}}_{\rm p}$ is the dense map predicted by a segmentation network $f(\theta): \textbf{\textit{X}}\mapsto{\textbf{\textit{Y}}_{\rm p}}$ with the $\theta$ as the parameters of the network $f$. $\textbf{\textit{Y}}_{\rm p}$ can be considered as a batch of pseudo label for unlabeled data $(\textbf{\textit{X}}_{\rm u}, \textbf{\textit{Y}}_{\rm p}) \in \mathbf{U}$ for retraining networks. Final evaluation results are calculated based on the differences between ${\textit{Y}}_{\rm p}$ and ${\textit{Y}}_{\rm gt}$ of $\mathbf{T}$. The training of S4CVnet framework is to minimize the sum of supervision loss ${\textit{Loss}}_{\rm sup}$ and the semi-supervision loss ${\textit{Loss}}_{\rm semi}$ which are based on the difference of inference of each network with ${\textit{Y}}_{\rm gt}$, and ${\textit{Y}}_{\rm p}$, respectively. There is no overlap between $\mathbf{L}$, $\mathbf{U}$ and $\mathbf{T}$ in our study. The framework of S4CVnet, as shown in Figure~\ref{fig:sscvnet}, consists of a feature-learning module and a guidance module which are based on three networks $f$, i.e. a CNN-based network $f_{\rm CNN}(\theta)$, and two ViT-based networks $f_{\rm ViT}(\theta)$. The $\theta$ of each network of the feature-learning module are initialized separately to encourage the difference of the two views of learning, and the $\theta$ of the guidance module is updated from one of the feature-learning networks which have the same architecture via EMA. The final inference of S4CVnet is considered as the output by guidance module $f_{\rm ViT}(\overline{\theta}): \textbf{\textit{X}} \mapsto {\textbf{\textit{Y}}}$. The details of CNN \& ViT networks, feature-learning module, and guidance module are discussed in the following Section \ref{cv}, \ref{flm}, and Section \ref{gm}, respectively.

\begin{figure*}[ht!]  
\centering  
\includegraphics[width=0.85\linewidth]{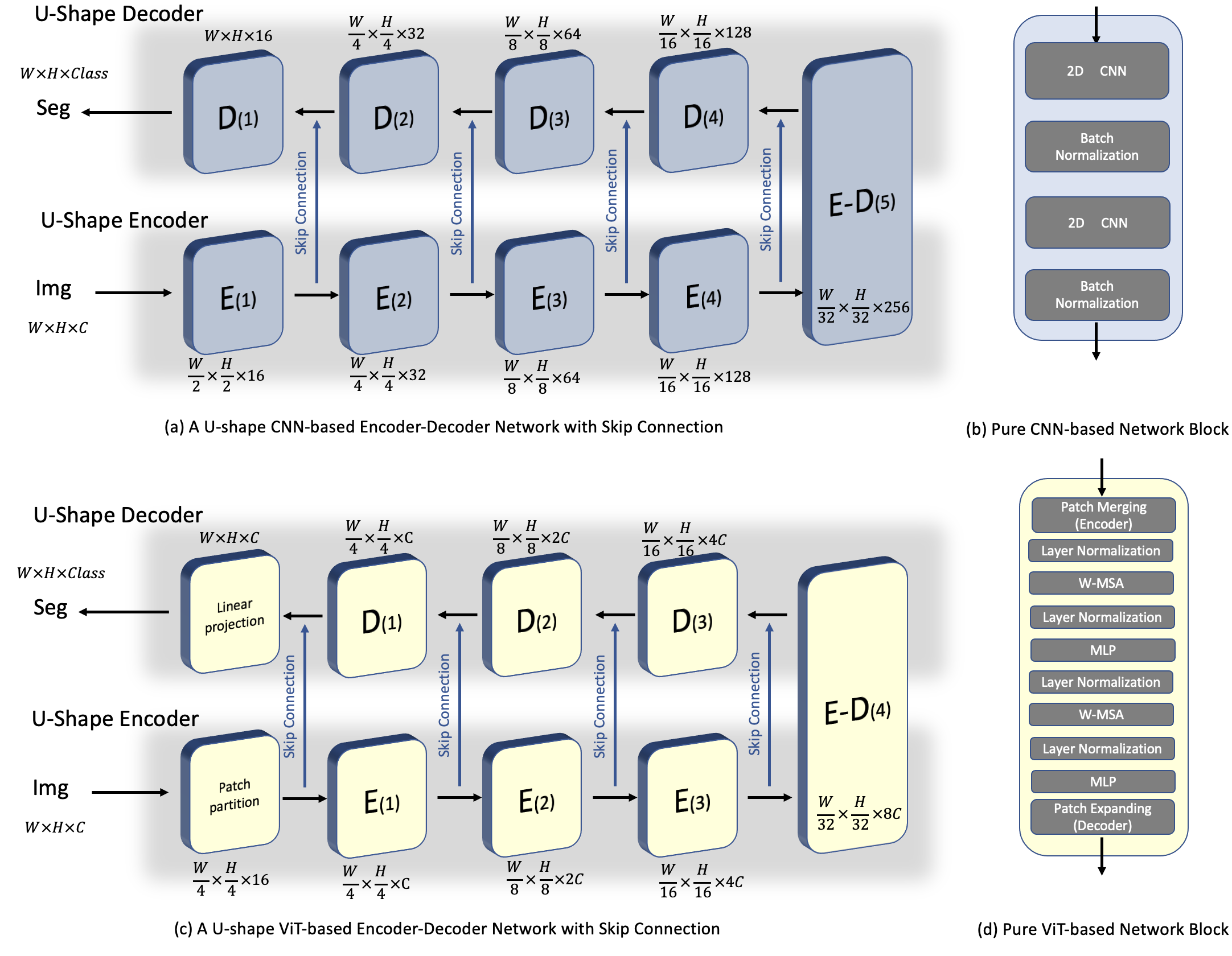}  
\caption{The Backbone Segmentation Network. (a,c)a U-shape CNN-based or ViT-based encoder-decoder style segmentation network, (b,d)a pure CNN-based or ViT-based network block. These two network blocks can be directly applied to the U-shape encoder-decoder network resulting in a purely CNN- or ViT-based segmentation network.}  
\label{fig:sscvnet2}  
\end{figure*}

\subsection{CNN \& ViT}\label{cv}
To fairly compare, analyse, and explore the feature learning ability of CNN and ViT, we propose a U-shape encoder-decoder style multi-class medical image semantic segmentation network, and it can be built with a purely CNN-based network block or ViT-based network block, respectively. Motivated by the success of the skip connection of U-Net~\cite{ronneberger2015u}, we firstly propose a U-shape segmentation network with 4 encoders and decoders connected by skip connections which are briefly sketched in Figure~\ref{fig:sscvnet2} (a). A pure CNN or ViT segmentation network hereby can be directly built with replacing the encoders and decoders with the proposed network blocks which is sketched in Figure~\ref{fig:sscvnet2} (b). In each of CNN-based block, two $3\times3$ convolutional layers and two batch normalization~\cite{ioffe2015batch} are developed accordingly~\cite{ronneberger2015u}. The ViT-based block is based on Swin-Transformer block~\cite{liu2021swin} with no further modification motivated by~\cite{chen2021transunet,cao2021swin}. Different with the traditional Transformer block~\cite{dosovitskiy2020image}, layer normalization ${\rm LN}$~\cite{ba2016layer}, multi-head self attention, residual connection~\cite{he2016deep}, ${\rm MLP}$ with GELU are developed with shift-window which results in window-based multi-head self attention(${\rm WMSA}$) and shifted window-based multi-head self attention(${\rm SWMSA}$). Both of WMSA and SWMSA are applied in the two successive transformer blocks respectively shown on the upside of Figure~\ref{fig:sscvnet2} (b). The details of data pipeline through self-attention-based ${\rm WMSA}$, ${\rm SWMSA}$, ${\rm MLP}$ for feature learning of ViT are summarised in Equations~\ref{equationa}, \ref{equationb}, \ref{equationc}, \ref{equationd}, and \ref{equatione}, where $i \in {1\cdots{L}} $, and $L$ is the number of blocks. The self-attention mechanism comprises three point-wise linear layers mapping tokens to intermediate representations: quires $\textbf{\textit{Q}}$, keys $\textbf{\textit{K}}$, and values $\textbf{\textit{V}}$, introduced in Equation~\ref{equatione}. In this way, the transformer block maps input sequence $\textbf{\textit{Z}}_0=[z_{0,1}\cdots{z}_{0,N}]$  positions to $\textbf{\textit{Z}}_L=[z_{L,1},...,z_{L,N}]$, and the much richer sufficient semantic feature information(global dependencies) is fully extracted and collected through the ViT-based block.

\small{
\begin{equation}
\textbf{\textit{Z}}_{i-1} = {\rm W\-MSA}({\rm LN} (\textbf{\textit{Z}}_{i-1})) + \textbf{\textit{Z}}_{i-1}
\label{equationa}
\end{equation}

\begin{equation}
\textbf{\textit{Z}}_{i} = {\rm MLP}({\rm LN}(\textbf{\textit{Z}}_{i})) + \textbf{\textit{Z}}_{i}
\label{equationb}
\end{equation}

\begin{equation}
\textbf{\textit{Z}}_{i+1} = {\rm SW\-MSA}({\rm LN} (\textbf{\textit{Z}}_{i})) + \textbf{\textit{Z}}_{i}
\label{equationc}
\end{equation}

\begin{equation}
\textbf{\textit{Z}}_{i+1} = {\rm MLP}({\rm LN}(\textbf{\textit{Z}}_{i+1})) + \textbf{\textit{Z}}_{i+1}
\label{equationd}
\end{equation}

\begin{equation}
{\rm MSA}(\textbf{\textit{Z}}') = {\rm softmax}(\frac{\textbf{\textit{Q}}\textbf{\textit{K}}}{\sqrt{D}})\textbf{\textit{V}}\\
\label{equatione}
\end{equation}
}
where $\textbf{\textit{Q}}, \textbf{\textit{K}}, \textbf{\textit{V}} \in \mathbb{R}^{M^{2} \times d}$, and $M^{2}$ represents the number of patches in a window, and $d$ is the dimension of the query and key.

Unlike conventional CNN-based blocks with downsampling and upsampling between each encoder or decoder, merging layers and expanding layers are designed between each ViT-based encoder, or decoder, respectively~\cite{chen2021transunet,cao2021swin}. The merging layer is designed to reduce 2 times of the number of tokens and increase 2 times of the feature dimension. It divides the input patches into 4 parts and concatenates them together. A linear layer is applied to unify the dimension to 2 times. The expanding layer is designed to reshape the size of input feature maps 2 times bigger, and reduces the feature dimension to half of the input feature map dimension. It uses a linear layer to increase the feature dimension, and then rearranges operation is used to expand the size and reduce the feature dimension to a quarter of the input dimension. A brief illustration of the size of the feature map in each step is in Figure~\ref{fig:sscvnet2} (a), where $W, H, C$ represents the width, height, and channel dimension of a feature map.  Considering making the ViT the same computational efficiency with the CNN for a fair comparison and complement each other, we come up with the setting: patch size is 4, input channel is 3, embedded dimension is 96, the number of head of self-attention is 3,6,12,24, the window size is 7, 2 swin-transformer-based blocks for each encoder/decoder, and the ViT is pre-trained with ImageNet~\cite{deng2009imagenet}. More details of CNN and ViT backbone with setting is available in Appendix.

\subsection{Feature-Learning Module}\label{flm}

The semi-supervised learning, especially in the pseudo-label-based approach, has been studied in image segmentation~\cite{blum1998combining,qiao2018deep,dong2018tri}. It incorporates segmentation inference on unlabeled data from one network as the pseudo label to retrain the other network, i.e. multi-view co-training approach~\cite{han2018co,luo2021semi}. Motivated by the recent success of cross pseudo label supervision~\cite{chen2021semi}, which has two networks $f(\theta_{1})$, $f(\theta_{2})$ with same architecture but initialized separately to encourage the difference of dual views, we further propose a feature-learning module aiming to explore the power of ViT and CNN mutually. Besides parameters of two networks are of course initialized separately, two completely different architectures of networks $f_{\rm CNN}(\theta_{1})$, $f_{\rm ViT}(\theta_{2})$ are designed to benefit each other via multi-view learning thus boost the performance of dual-view learning. 
The proposed feature-learning module to generate pseudo label can be illustrated as:
\begin{equation}
P_1 = f_{\rm CNN}(X;\theta_1), P_2 = f_{\rm ViT}(X;\theta_2).
\end{equation}
where $\theta_1, \theta_2$ demonstrate network are initialized separately, $P_1, P_2$ represent the segmentation inference with $f_{\rm CNN}(\theta_{1})$, $f_{\rm ViT}(\theta_{2})$, respectively. The pseudo label based on $P_1, P_2$ then is utilized to supervise and complement each other. The CNN is mainly based on the local convolution operation, but the ViT is to model the global dependencies of feature through self-attention~\cite{dosovitskiy2020image}, so two segmentation inferences $P_1, P_2$ have different properties of prediction, and no explicit constraints to enforce two inferences similar. The supervision detail of simultaneously complementing each other (update parameters of ViT and CNN) is discussed in Section~\ref{objectives}.

\subsection{Guidance Module\label{gm}}
Except for the feature-learning module to enable two networks to learn from the data, a robust guidance module is designed under the consistency-aware concern to boost the performance and also act as the final module for evaluation of S4CVnet. Inspired by temporal ensembling~\cite{laine2016temporal}, and self-ensembling~\cite{tarvainen2017mean}, the guidance network is to further supervise the perturbed networks and minimize the inconsistency. In a training process, the perturbation is firstly applied to the one of a network in the feature-learning module. Secondly, the parameter of the network is updated iteratively with back prorogation. Then the network of guidance module is updated via exponential moving average(EMA) from the feature-learning module. Finally, a much more robust guidance module which is more likely to be correct than the feature learning network is then to supervise two feature-learning networks with the consistency concern. In S4CVnet, the guidance module is based on ViT which has the same architecture as the ViT in the feature-learning network, so that guidance ViT can be constantly updated through EMA of the parameter of the ViT network learning from the data~\cite{laine2016temporal}. 
The proposed guidance module to generate pseudo label can be illustrated as:
\begin{equation}
P_3 = f_{\rm ViT}(X;\overline{\theta}).
\end{equation}
where $\overline{\theta}$ demonstrates the network ViT is based on averaging network weights rather than directly trained by the data. $\overline{\theta}$ is updated based on the parameter of feature-learning ViT model $\theta_{t}$ on past training step $t$, which can be illustrated as $\overline{\theta} = \alpha \theta_{t-1} + (1-\alpha)\theta_{t}$. $\alpha$ is a weight factor which is calculated as the $\alpha=1-\frac{1}{t+1}$. $P_3$ represents the segmentation inference with $f_{\rm ViT}(\overline{\theta})$, which is used to supervise the feature-learning module following the consistency-aware concern. The supervision details of feature-learning ViT and CNN by guidance module are discussed in Section~\ref{objectives}.

\subsection{Objective}\label{objectives}
The training objective is to minimize the sum of the supervision loss $\mathcal{L}_{\rm sup}$ and the semi-supervision $\mathcal{L}_{\rm semi}$ among the three networks $f_{\rm CNN}(\theta_{1})$,$f_{\rm ViT}(\theta_{2})$, and $f_{\rm ViT}(\overline{\theta})$, so the overall loss of S4CVnet being optimized during training is detailed in Equation \ref{equation1}:
\begin{equation}
\begin{aligned}
\mathcal{L} = \mathcal{L}_{\rm sup 1} + \mathcal{L}_{\rm sup 2} + \lambda_{1} (\mathcal{L}_{\rm semi 1}+\mathcal{L}_{\rm semi 2})+
\lambda_{2}(\mathcal{L}_{\rm semi 3}+\mathcal{L}_{\rm semi 4} )
\label{equation1}
\end{aligned}
\end{equation}
where $\lambda_1, \lambda_2$ are the weight factor of cross-supervision dual-view loss and consistency-aware loss, and it is updated every 150 iterations~\cite{laine2016temporal}. It is a trade-off weight that keeps increasing during the training process to make S4CVnet focus on labelled data when initialize, and then move focus to unlabeled data with our proposed semi-supervision approach. This is made under the assumption of the S4CVnet can gradually infer much reliable pseudo label confidently. The weight factor is briefly indicated in Equation \ref{equationlambda}.

\begin{equation}
\lambda = e^{-5\times(1-t_{\rm iteration}/t_{\rm maxiteration})^{2}}
\label{equationlambda}
\end{equation}
where $t$ indicates the current iteration number in a complete training process. Each of $\mathcal{L}_{\rm sup}$ and $\mathcal{L}_{\rm semi}$ are discussed as follows:

The semi-supervision loss among each network $\mathcal{L}_{\rm semi}$ are calculated based on Cross-Entropy $\mathrm{CE}$ as shown in Equation \ref{equation2}:
\begin{equation}
\begin{aligned}
\mathcal{L}_{\rm semi} = \mathrm{CE}\big({\rm argmax}{(f_{1}( {\textbf{\textit{X}}}; \theta), f_{2}( {\textbf{\textit{X}}}; \theta) )}  \big)\\
\label{equation2}
\end{aligned}
\end{equation}
here we simply four $\mathcal{L}_{\rm semi}$ losses with a pair of $\big(f_{1}( {\textbf{\textit{X}}}; \theta), f_{2}( {\textbf{\textit{X}}}; \theta) \big)$, where the pair can be $\big(f_{\rm CNN}( {\textbf{\textit{X}}}; \theta_{\rm 1}), f_{\rm ViT}( {\textbf{\textit{X}}}; \theta_{\rm 2}) \big)$, $\big(f_{\rm ViT}( {\textbf{\textit{X}}}; \theta_{\rm 2}), f_{\rm CNN} {\textbf{\textit{X}}}; \theta_{\rm 1}) \big)$, 
$\big(f_{\rm ViT}( {\textbf{\textit{X}}}; \overline{\theta}), f_{\rm ViT}( {\textbf{\textit{X}}}; \theta_{\rm 2}) \big)$, and $\big(f_{\rm ViT}( {\textbf{\textit{X}}}; \overline{\theta}), f_{\rm CNN}( {\textbf{\textit{X}}}; \theta_{\rm 1}) \big)$.

The supervision loss $\mathcal{L}_{\rm sup}$ for each network is calculated based on both $\mathrm{CE}$ and the Dice Coefficient $\mathrm{Dice}$ as shown in  Equation \ref{equation3}:
\begin{equation}
\begin{aligned}
\mathcal{L}_{\rm sup} = \frac{1}{2} \times \big(  \mathrm{CE}({\textit{Y}}_{\rm gt}, f( {\textbf{\textit{X}}}; \theta) ) +\mathrm{Dice}({\textit{Y}}_{\rm gt}, f( {\textbf{\textit{X}}}; \theta))\big)
\label{equation3}
\end{aligned}
\end{equation}
Here we simply two $\mathcal{L}_{\rm sup}$ supervision losses with a network $f( {\textbf{\textit{X}}}; \theta)$, which can be considered as $f_{\rm CNN}(\theta_{\rm 1})$, and $f_{\rm ViT}(\theta_{\rm 2})$, because each network trained with labeled data $ \textit{Y}_{\rm gt}$ set is directly with the same way. The S4CVnet and all other baseline methods reported in Section \ref{baseline} are with the same loss design including $\mathrm{CE}$, and $\mathrm{Dice}$ for $\mathcal{L}_{\rm sup}$ and $\mathcal{L}_{\rm semi}$ in order to conduct a fair comparison.

\section{Experiments and Results}\label{baseline}
\subsubsection{Data set}
Our experiments validate the S4CVnet and all other baseline methods on the MRI ventricle segmentation data set from the automated cardiac diagnosis MICCAI Challenge 2017~\cite{bernard2018deep}. The data  is from 100 patients (nearly 6\,000 images) covering different distributions of feature information, across five evenly distributed subgroups: normal, myocardial infarction, dilated cardiomyopathy, hypertrophic cardiomyopathy, and abnormal right ventricle. All images are resized to $224{\times}224$. 20\% of images are selected as the testing set, and the rest of the data set is for training(including validation). 

\subsubsection{Implementation Details}
Our code has been developed under Ubuntu 20.04 in Python 3.8.8 using Pytorch 1.10 and CUDA 11.3 using four Nvidia GeForce RTX 3090 GPU, and Intel(R) Intel Core i9-10900K. The runtimes averaged around 5 hours, including the data transfer, training, inference and evaluation. The data set is processed for 2D image segmentation purposes. S4CVnet is trained for 30,000 iterations, the batch size is set to 24, the optimizer is SGD, and the learning rate is initially set to 0.01, momentum is 0.9, and weight decay is 0.0001. The network weight is saved and evaluated on the validation set every 200 iterations, and the network of guidance module with the best validation performance is used for final testing. The setting is also applied to other baseline methods directly without any modification.

\subsubsection{Backbone}
The S4CVnet consists of two types of networks as shown in Figure \ref{fig:sscvnet}. One is CNN-based segmentation network with skip connection, UNet\cite{ronneberger2015u}, and the other one is ViT-based segmentation network with shift window\cite{liu2021swin} and skip connection, Swin-UNet\cite{cao2021swin}. For a fair comparison, two networks are both with U-shaped architecture with purely CNN- or ViT-based blocks as encoders and decoders. The tiny version of ViT is selected in this study to make the computational cost and training efficiency similar to CNN.

\subsubsection{Baseline Methods}
All methods including S4CVnet and other baseline methods are trained with the same hyper-parameter setting, and the same distribution of features. The randomly selection of test set, labelled train set and unlabeled train set are only conducted once and then tested with all baseline methods together as well as S4CVnet. The baseline methods reported includes: MT~\cite{tarvainen2017mean}, DAN~\cite{zhang2017deep}, ICT~\cite{verma2019interpolation},  ADVENT~\cite{vu2019advent}, UAMT~\cite{yu2019uncertainty}, DCN~\cite{qiao2018deep}, CTCT~\cite{luo2021semi} with CNN as the backbone segmentation network.

\subsubsection{Evaluation Measures}
The direct comparison experiments between S4CVnet and other baseline methods are conducted with a variety of evaluation metrics including similarity measures: Dice, IOU, Accuracy, Precision, Sensitivity, and Specificity, which are the higher the better. The difference measures are also investigated: Hausdorff Distance (HD), and Average Surface Distance (ASD), which are the lower the better. The mean value of these metrics is reported, because the data set is a multi-class segmentation data set. The full evaluation measures are reported when comparing S4CVnet against other baseline methods, and the topological exploration of all alternative frameworks. IOU as the most common metric is also selected to report the performance of all baseline methods and S4CVnet under the assumption of different ratios of labelled data/total data. IOU, Sensitivity, and Specificity are selected to report the ablation study of different networks with different combinations of our proposed contribution. 

\subsubsection{Qualitative Results}
\begin{figure*}
\centering  
\includegraphics[width=\linewidth]{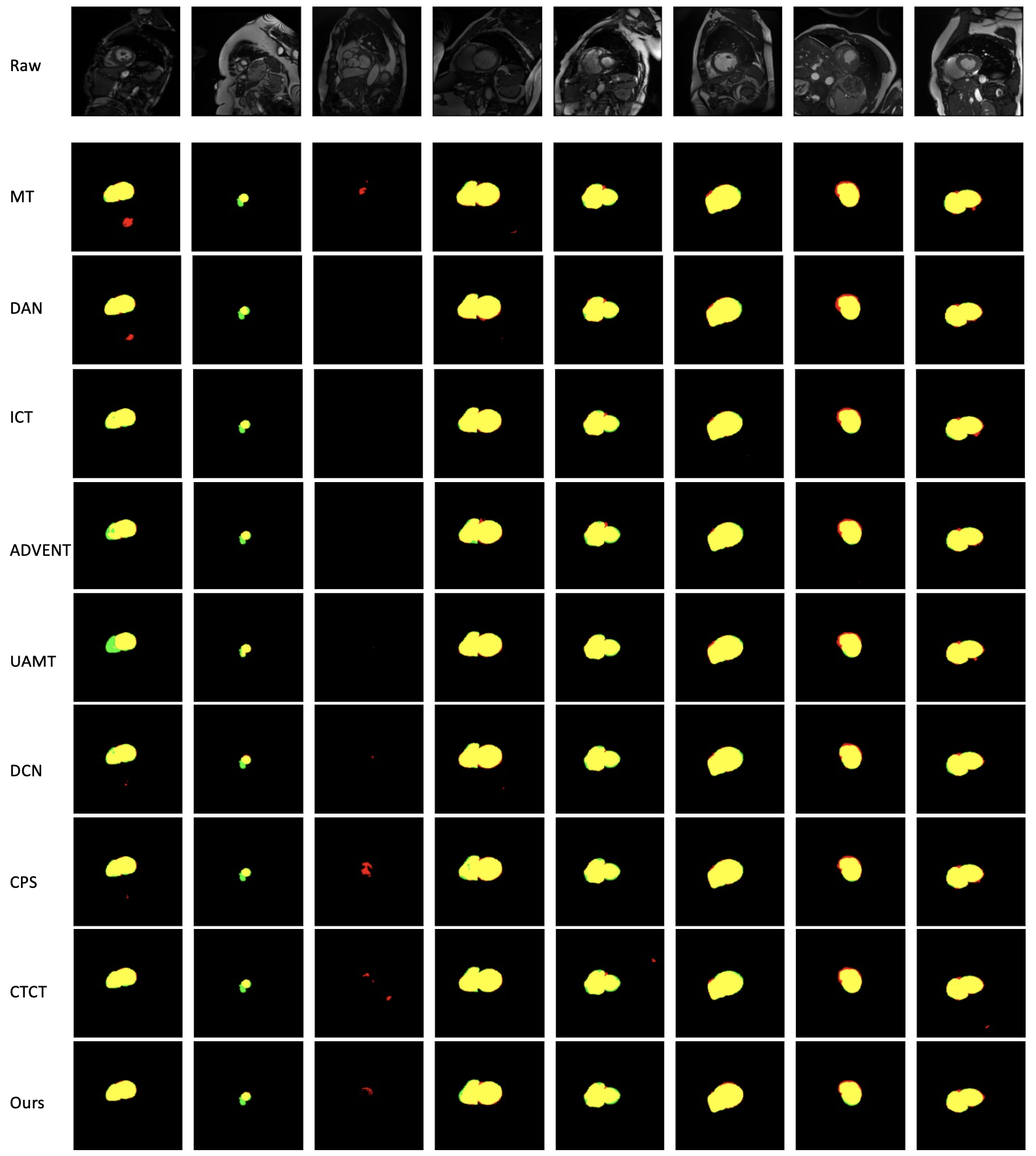}  
\caption{Sample Qualitative Results on MRI Cardiac Test Set. Yellow, Red, Green, and Black Indicate True Positive, False Positive, False Negative, and True Negative of Each Pixel.}

\label{fig:resultsimage}  
\end{figure*}

Figure~\ref{fig:resultsimage} illustrates eight randomly selected sample raw images with related predicted images against the published ground truth, where \textcolor{yellow}{Yellow}, \textcolor{red}{Red}, \textcolor{green}{Green} and Black represent as True Positive(TP), False Positive(FP), False Negative(FN) and True Negative(TN) inferences at pixel level, respectively. This illustrates how S4CVnet can give rise to fewer FP pixels and lower ASD compared to other methods.

\subsubsection{Quantitative Results}
Table \ref{tablebaseline} reports the direct comparison of S4CVnet against other semi-supervised methods including similarity measures and difference measures when the ratio of assumed labelled data/total data is 10\%. The best result of different measures on the table is in {\bf Bold}.

\

\begin{table*}[htbp]

\centering
\begin{tabular}{c|cccccc|cc}
\hline
Framework & mDice$\uparrow$ & mIOU$\uparrow$ & Acc$\uparrow$ & Pre$\uparrow$ & Sen$\uparrow$ & Spe$\uparrow$ & HD$\downarrow$ & ASD$\downarrow$  \\
\hline

MT\cite{tarvainen2017mean} & 0.8860 & 0.8034 & 0.9952 & 0.8898 & 0.8829 & 0.9720 & 9.3659 & 2.5960 \\

DAN\cite{zhang2017deep} & 0.8773 & 0.7906 & 0.9947 & 0.8721 & 0.8832 & 0.9743 & 9.3203 & 3.0326\\

ICT\cite{verma2019interpolation} & 0.8902 & 0.8096 & 0.9954 & 0.8916 & 0.8897 & 0.9745 & 11.6224 & 3.0885\\

ADVENT\cite{vu2019advent} & 0.8728 & 0.7836 & 0.9947 & 0.8985 & 0.8517 & 0.9601 & 9.3203 & 3.5026 \\
 
UAMT\cite{yu2019uncertainty} & 0.8683 & 0.7770 & 0.9946 & 0.8988 & 0.8416 & 0.9582 & 8.3944 & 2.2659 \\

DCN\cite{qiao2018deep} & 0.8809 & 0.7953 & 0.9951 & 0.8915 & 0.8714 & 0.9690 & 8.9155 & 2.7179 \\

tri\cite{chen2021semi} & 0.8918 & 0.7906 & 0.9947 & 0.8721 & 0.8832 & 0.9743 & {\bf 7.2026} & 2.2816 \\
\hline
2-Model-Based Group \cite{luo2021semi} & 0.8998 & 0.8245 & 0.9959 & 0.8920 & 0.9083 & 0.9825 & 9.6960 & 2.7293\\

{\bf S4CVnet, 3-Model}  &  {\bf 0.9146} & {\bf 0.8478} & {\bf 0.9966} & {\bf 0.9036} &  {\bf 0.9283} &  {\bf 0.9881} &  12.5359 & {\bf 0.6934} \\

4-Model-Based Group  & 0.8520 & 0.7555 & 0.9944 & 0.8654 & 0.8396 & 0.9672& 22.8588 & 2.1686\\

5-Model-Based Group  & 0.8714 & 0.7832 & 0.9951 & 0.8943 & 0.8507 & 0.9667 & 18.2431 & 1.6345 \\
\hline
\end{tabular}
\caption{Direct Comparison of Semi-supervised Frameworks on MRI Cardiac Test Set}
\label{tablebaseline}
\end{table*}

\begin{figure*}
\centering  
\includegraphics[width=\linewidth]{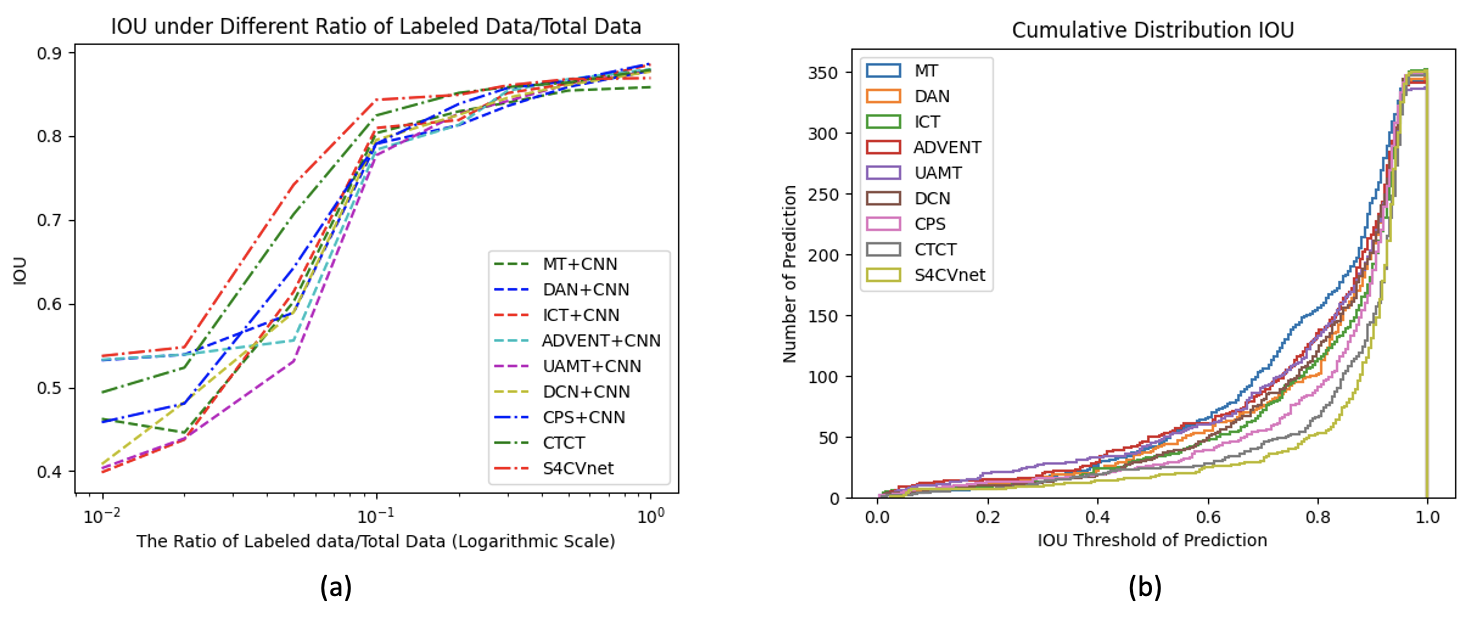}  
\caption{The Performance of S4CVnet Against Other Baseline Methods. (a) The line chart of mIOU results on the test set with different assumptions of the ratio of label/total data for training. (b) The histogram chart indicates the cumulative distribution of IOU performance of the predicted image on the test set.}

\label{fig:resultsimagelinechart}  
\end{figure*}

\begin{table}[t]
\begin{center}
\centering
\begin{tabular}{ cc|c|c|c|c|c }
\hline
\multicolumn{2}{c|}{Learning Module} & Guidance Module & \multirow{2}{*}{Test Network}& \multirow{2}{*}{IOU$\uparrow$} & \multirow{2}{*}{Sen$\uparrow$} & \multirow{2}{*}{Spe$\uparrow$} \\

Network A & Network B & Network C & &  & &  \\

 \hline
ViT & ViT&  \xmark & A & 0.8034 & 0.8829 & 0.9720 \\
ViT & ViT&  \xmark & B & 0.8135 & 0.9036 & 0.9821  \\
CNN & CNN&  \xmark & A & 0.7906 & 0.8832 & 0.9743 \\
CNN & CNN&  \xmark  & B & 0.8231 & 0.8967 & 0.9761 \\
\xmark & CNN& CNN  & B & 0.7345 & 0.8094 & 0.9586 \\
\xmark & CNN& CNN  & C & 0.7660 & 0.8481 & 0.9585 \\
\xmark & ViT& ViT  & B & 0.8159 & 0.9032 & 0.9822 \\
\xmark & ViT& ViT  & C & 0.7359 & 0.8415 & 0.9716\\
ViT & ViT& ViT  & A & 0.8096 & 0.8995 & 0.9817 \\
ViT & ViT& ViT  & B & 0.8194 & 0.9078 & 0.9833 \\
ViT & ViT& ViT  & C & 0.8183 & 0.9037 & 0.9822 \\
CNN & CNN& CNN  & A & 0.8399 & 0.9225 & 0.9848 \\
CNN & CNN& CNN  & B & 0.8432 & 0.9189 & 0.9848 \\
CNN & CNN& CNN  & C & 0.8345 & 0.9168 & 0.9828 \\
CNN & ViT& ViT  & A & 0.8341 & 0.9135 & 0.9825 \\
CNN & ViT& ViT  & B & 0.8354 & 0.9177 & 0.9839 \\
CNN & ViT& ViT  & C & {\bf 0.8478} &{\bf  0.9283} & {\bf 0.9881 }\\

 \hline
\end{tabular}
\caption{Ablation Studies on Contributions of Architecture and Modules}
\label{tab:ablationcomparison}
\end{center}
\end{table}

\subsubsection{Ablation Study}
In order to analyze the effects of each of the proposed contributions and combinations including the setting of the network, the mechanism of the feature-learning module and guidance module, and the robustness of each network of S4CVnet, extensive ablation experiments have been conducted and reported in Table~\ref{tab:ablationcomparison}. \xmark indicates either a network of feature-learning module or a guidance module is removed, and all alternative network settings(CNN or ViT) are explored. In different combinations of the proposed contribution, all the available networks are also tested separately, and the ablation study demonstrates that the proposed S4CVnet is with the most proper setting to fully utilize the power of CNN and ViT via student-teacher guidance scheme and dual-view co-training feature-learning approach in semi-supervised image semantic segmentation.

\begin{figure*}
\centering  
\includegraphics[width=0.9\linewidth]{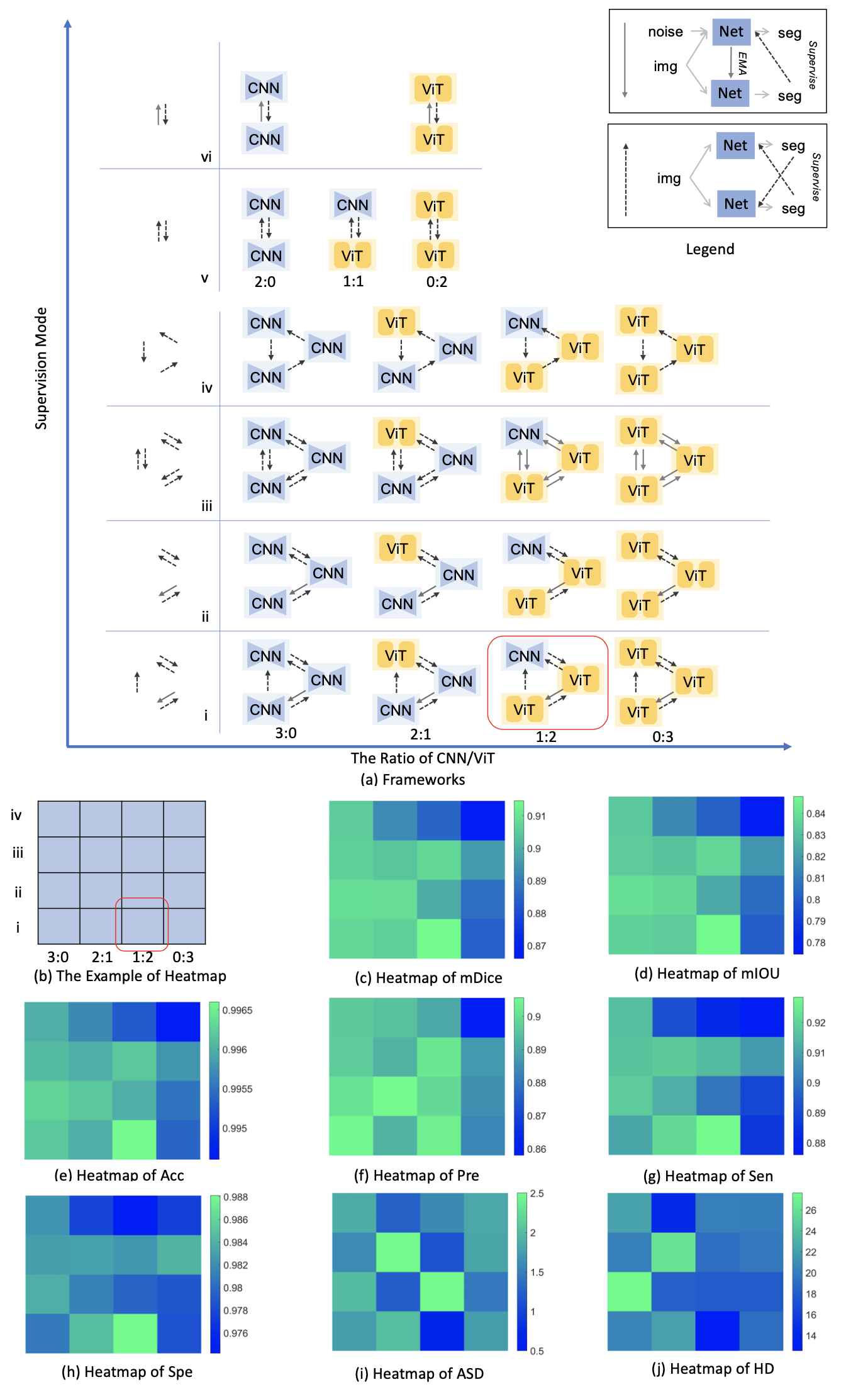}  
\caption{The Topological Exploration of the Network(CNN\&ViT), and  Semi-Supervised Supervision Mode (Student-Teacher Style \& Pseudo-Label).}

\label{fig:ablationtopological}  
\end{figure*}

\subsubsection{Supervision Mode Exploration}
Besides the ablation study to explore the different settings and combination of networks, feature-learning module, and guidance module, we fully explore the semi-supervised learning in medical image semantic segmentation through topological exploration of all alternative supervision modes of CNN and ViT. The full list of alternative frameworks is illustrated in Figure~\ref{fig:ablationtopological}, where two supervision mode is briefly sketched in the legend of the figure. $\longrightarrow$ indicates the Student-Teacher style supervision mode, and $\dashrightarrow$ indicate cross pseudo-label-based supervision mode. Figure~\ref{fig:ablationtopological} (a) briefly illustrates all alternative frameworks with two axes, the Y-axis with different supervision modes from three networks to two networks, and the X-axis with the ratio of the number of CNN/ViT networks, and  the proposed S4CVnet is in a \textcolor{red}{red} bounding box. All frameworks shown in the Figure~\ref{fig:ablationtopological} (a) have been tested and reported with heatmap format directly. Figure~\ref{fig:ablationtopological} (b) is an example heatmap to indicate the supervision mode and ratio of CNN/ViT information depending on the position of heatmap with a \textcolor{red}{red} bounding box to illustrate where is the S4CVnet as well.  Figure~\ref{fig:ablationtopological} (c,d,e,f,g,h,i,j) represent the heatmap with mDice, mIOU, accuracy, precision, sensitivity, specificity, ASD, and HD validation performance, which demonstrate a whole picture of semi-supervised learning for medical semantic segmentation with CNN and ViT, and the denominating position of our proposed S4CVnet. The details of the quantitative results of topological exploration is in Appendix.

\section{Conclusions}
In this paper, we introduce an advanced semi-supervised learning framework in medical image semantic segmentation, S4CVnet, aiming to fully utilize the power of CNN and ViT simultaneously. S4CVnet consists of a feature-learning module and a guidance module. The feature-learning module, a dual-view feature learning approach, is proposed to enable two networks to complement each other via pseudo-label supervision. The guidance module is based on averaging network weights to supervise the learning modules under the consistency concern. Our proposed methods is evaluated with a variety of evaluation metrics and different assumption of the ratio of labelled data/total data against other semi-supervised learning baselines with the same hyperparameters settings and keeps the state-of-the-art position on a public benchmark data set. Besides a comprehensive ablation study, a topological exploration with CNN and ViT illustrates a whole picture of utilizing CNN and ViT in semi-supervised learning.

\clearpage

\bibliographystyle{splncs04}
\bibliography{egbib}


\section{Appendix / Supplementary Material}

\subsection{Details of Experiments and Code}

We used UNet and Swin-UNet as the segmentation backbone of CNN and ViT, respectively. All the baseline methods, proposed S4CVnet, and topological exploration of semi-supervised learning with CNN and ViT are developed with the same hyperparameter setting including optimizer, learning rate, batch size, and loss function. The feature distribution of labelled data set, unlabeled data set, validation data set, and the test data set is same for all methods in the experiment section. \\
The ViT backbone network is available at \footnote[2]{https://github.com/HuCaoFighting/Swin-Unet}, all baseline methods is available at \footnote[3]{https://github.com/HiLab-git/SSL4MIS} without any modification. To reproduce these results, S4CVnet code is available at \footnote[4]{https://github.com/ziyangwang007/CV-SSL-MIS}.\\

\subsection{Details of Topological Exploration Results}
\begin{figure*}
\centering  
\includegraphics[width=0.9\linewidth]{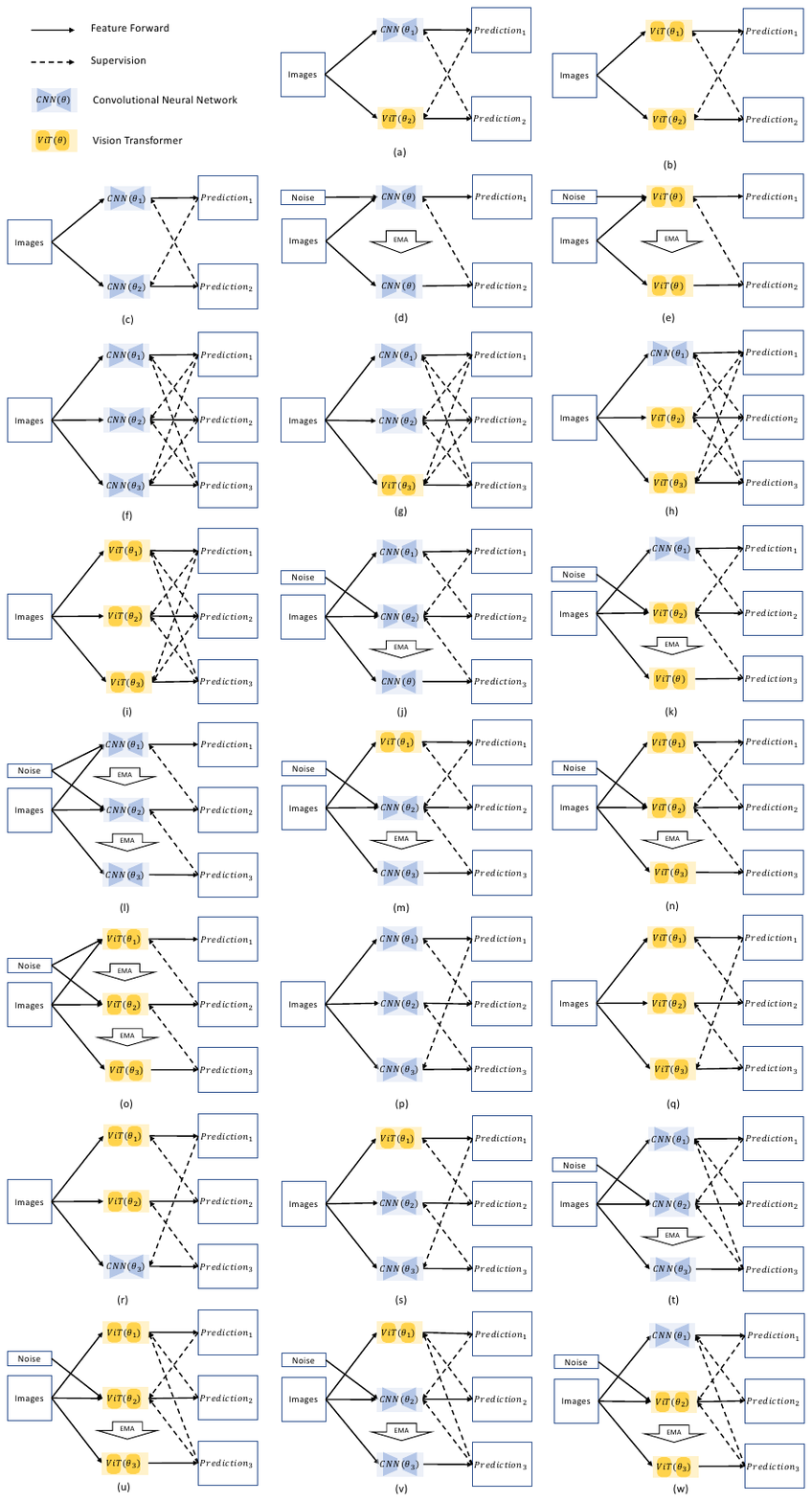}  
\caption{The Full List of All Alternative Supervision Modes with CNN \& ViT Frameworks with Our Proposed Techniques}

\label{fig:ablationappendix}  
\end{figure*}

In Section~\ref{baseline}, a topological exploration with supervision modes and networks for all alternative semi-supervision frameworks in image semantic segmentation has been tested, and reported with heatmaps to give a whole picture of proposed methods. To clearly illustrate our topological exploration, all frameworks are simply sketched in Figure~\ref{fig:ablationappendix}, and the full quantitative results of all frameworks are reported in Table~\ref{tablebaselineappendix} accordingly.

\begin{table*}[htbp]

\centering
\begin{tabular}{c|cccccc|cc}
\hline
Supervision Mode & mDice$\uparrow$ & mIOU$\uparrow$ & Acc$\uparrow$ & Pre$\uparrow$ & Sen$\uparrow$ & Spe$\uparrow$ & HD$\downarrow$ & ASD$\downarrow$  \\
\hline

A & 0.8998 & 0.8245 & 0.9959 & 0.8920 & 0.9083 & 0.9825 & 9.6960 & 2.7293 \\
B & 0.8927 & 0.8135 & 0.9956 & 0.8832 & 0.9036 & 0.9821 & 17.7406& 1.6316 \\
C & 0.8918 & 0.7906 & 0.9947 & 0.8721 & 0.8832 & 0.9743 & {\bf 7.2026} & 2.2816 \\
D & 0.8860 & 0.8034 & 0.9952 & 0.8898 & 0.8829 & 0.9720 & 9.3659 & 2.5960 \\
E & 0.8384 & 0.7359 & 0.9938 & 0.8361 & 0.8415 & 0.9716 & 23.7689 & 2.2801 \\
F & 0.9061 & 0.8341 & 0.9961 & 0.8970 & 0.9165 & 0.9829 & 20.1008 & 1.6110 \\
G & 0.9042 & 0.8311 & 0.9960 & 0.8918 & 0.9182 & 0.9831 & 26.2525 & 3.1882 \\
H & 0.9077 & 0.8372 & 0.9962 & 0.9022 & 0.9149 & 0.9825 & 19.1385 & 1.1296 \\
I & 0.8958 & 0.8184 & 0.9958 & 0.8866 & 0.9084 & 0.9841 & 19.7125 & 1.8150 \\
J & 0.9092 & 0.8391 & 0.9963 & 0.9006 & 0.9186 & 0.9838 & 27.6241 & 1.9961 \\
K & 0.8995 & 0.8243 & 0.9960 & 0.8993 & 0.9004 & 0.9797 & 17.9150 & 4.4550 \\
M & 0.9084 & 0.8380 & 0.9962 &{\bf  0.9056 }& 0.9125 & 0.9814 & 18.2136 & 1.2208 \\
N & 0.8872 & 0.8054 & 0.9955 & 0.8856 & 0.8898 & 0.9788 & 17.9167 & 1.4451 \\
P & 0.9054 & 0.8330 & 0.9960 & 0.8965 & 0.9153 & 0.9823 & 22.2934 & 1.8733 \\
Q & 0.8660 & 0.7744 & 0.9946 & 0.8580 & 0.8760 & 0.9774 & 20.0766 & 1.8373 \\
R & 0.8854 & 0.8030 & 0.9953 & 0.8900 & 0.8819 & 0.9742 & 20.1741 & 1.5861 \\
S & 0.8930 & 0.8141 & 0.9957 & 0.8951 & 0.8919 & 0.9778 & 14.2738 & 1.2712 \\
T & 0.9074 & 0.8359 & 0.9962 & 0.9051 & 0.9104 & 0.9809 & 20.4208 & 1.7543 \\
U & 0.8843 & 0.8010 & 0.9954 & 0.8834 & 0.8860 & 0.9786 & 19.1564 & 1.7462 \\
V & 0.9060 & 0.8339 & 0.9960 & 0.8921 & 0.9212 & 0.9847 & 22.1800 & 2.0180 \\

\hline
{\bf W(Ours)}  & {\bf 0.9146} &{\bf 0.8478} & {\bf 0.9966} & 0.9036 &{\bf 0.9283} & {\bf 0.9881} & 12.5359 & {\bf 0.6934} \\
\hline
\end{tabular}
\caption{Direct Comparison of All Alternative Semi-supervised Frameworks with Our Proposed Techniques}
\label{tablebaselineappendix}
\end{table*}

\subsection{Details of Qualitative Results Under Assumption of Different Rate of Labeled/Total Data}

Table \ref{tab:ablationdicecomparisonv2} reports mIOU of S4CVnet and other baseline methods on test set under the different assumption of ratio of labeled data/total data for training, i.e. 1\%, 2\%, 5\%, 10\%, 20\%, 30\%, 50\%, and 100\%, respectively . 

\begin{table}[]

\centering
\begin{tabular}{c|ccccccccc}
\hline
Labeled/Total & 1\% & 2\%  & 5\% & 10\%  & 20\% &30\% & 50\% & 100\% \\
\hline

MT\cite{tarvainen2017mean} & 0.4623  & 0.4460 & 0.6021 & 0.8034 &  0.8294 & 0.8397 & 0.8542 & 0.8583 \\
DAN\cite{zhang2017deep}  &  0.5323 & 0.5391 & 0.5892 & 0.7906 & 0.8130 & 0.8356 & 0.8585 & 0.8780\\
ICT\cite{verma2019interpolation} &  0.3985 & 0.4376 & 0.6140 & 0.8096 & 0.8191& 0.8512 &0.8624 & 0.8853\\
ADVENT\cite{vu2019advent} &  0.5329  & 0.5391 & 0.5559 & 0.7836 & 0.8133 &0.8537 &0.8677 &0.8797 \\
UAMT\cite{yu2019uncertainty}   & 0.4034 & 0.4390 & 0.5310 & 0.7770 & 0.8269 & 0.8416 & 0.8619 & 0.8778\\
DCN\cite{qiao2018deep} & 0.4083 & 0.4824 & 0.5896 & 0.7953 &  0.8252 & 0.8455 & 0.8610 & 0.8769\\
CPS\cite{chen2021semi} & 0.4583 & 0.4806 & 0.6426 & 0.7906 &   0.8383& 0.8572 &0.8654  & {\bf 0.8865}\\
CTCT\cite{luo2021semi} & 0.4939 & 0.5235 & 0.7066 & 0.8245 &{\bf 0.8515} & 0.8584 & 0.8638 & 0.8791\\
\hline
{\bf S4CVnet}  & {\bf 0.5374} & {\bf 0.5479} &{\bf 0.7418} &{\bf 0.8432} & 0.8491 & {\bf 0.8604} & {\bf 0.8679} & 0.8691 \\

\hline
\end{tabular}
\caption{The Mean IOU Results on Test Set Under Different Assumptions of Ratio of Label/Total Data for Training}
\label{tab:ablationdicecomparisonv2}
\end{table}




\end{document}